\documentclass[12pt]{iopart}
\usepackage{graphicx,color}
\usepackage{bm}
\usepackage{amssymb}
\newcommand{\text}{\mathrm}
\newcommand\bea{\begin{eqnarray}}
\newcommand\eea{\end{eqnarray}}
\newcommand\nn{\nonumber}
\newcommand\la{\langle}
\newcommand\ra{\rangle}
\begin{document}
\title{Large deviations of heat flow in harmonic chains}

\author{Anupam Kundu, Sanjib Sabhapandit, and Abhishek Dhar}
\address{Raman Research Institute, Bangalore 560080, India}


\begin{abstract}
We consider heat transport across a harmonic chain connected at its
two ends to white-noise Langevin reservoirs at different
temperatures. In the steady state of this system the heat $Q$ flowing
from one reservoir into the system in a finite time $\tau$ has a
distribution $P(Q,\tau)$. We study the large time form of the
corresponding moment generating function $\langle \rme^{-\lambda
  Q}\rangle\sim g(\lambda)~\rme^{ \tau\mu (\lambda)}$. Exact formal
expressions, in terms of phonon Green's functions , are obtained for
both $\mu(\lambda)$ and also the lowest order correction $g(\lambda)$.
We point out that, in general a knowledge of both $\mu(\lambda)$ and
$g(\lambda)$ is required for finding the large deviation function
associated with $P(Q,\tau)$.  The function $\mu(\lambda)$ is known to
be the largest eigenvector of an appropriate Fokker-Planck type
operator and our method also gives the corresponding eigenvector
exactly.
\end{abstract}

\maketitle

\section{Introduction}
Among the most interesting recent developments in the theory of
nonequilibrium processes are the so-called fluctuation theorems
\cite{ECM93,ES94,GC95,Jarz97,Kur98,Crooks99,LS99}. These theorems make
quantitative statements on the probability of negative entropy
production in nonequilibrium systems. They have been theoretically
\cite{HS01,ND04,Seifert05,JW04,BD04,ED04,DDR04,DL98,visco06,keijidhar,AG06,lacoste08}
and experimentally
\cite{Wang02,Carberry04,FM04,Goldburg01,Douarche06,Liphardt02,Collin05,Solano10,MS08}
studied in a large number of systems in various nonequilibrium
states. The results have been obtained both in the context of
transient and steady state phenomena.  In the case of nonequilibrium
steady states in systems carrying heat or particle current, the
fluctuation theorems have pointed to the importance of the large
deviation function (LDF) and the cumulant generating function (CGF)
\cite{touchette:2009}. The steady state fluctuation theorem can, in
these cases, be equivalently stated as a symmetry property of the LDF
or the CGF \cite{BD04,DDR04,DL98}.  Apart from their interest from the
point of the fluctuation theorem these two functions contain important
information on nonequilibrium processes: the LDF gives the precise
probability of the occurrence of rare events while the CGF contains
information on the average current in a system as well as all moments
\cite{touchette:2009}.  So far there are very few examples where
either the LDF or the CGF have been exactly computed. The few examples
include particle transport in exclusion processes
\cite{BD04,DDR04,DL98}, Brownian motors \cite{AG06,lacoste08}, power
dissipation and heat transport in single Brownian particles
\cite{visco06,gomez,Wijland:2006,farago:2002} and heat conduction
across a quantum harmonic chain \cite{keijidhar}.

Consider the example of heat conduction
through a system coupled to two heat baths at different temperatures
$T_L$ and $T_R$ and let $Q$ be the heat flowing from the left
reservoir into the system during a time interval $\tau$. Then $Q$ is a
stochastic variable with a distribution $P(Q)$ and the LDF and CGF for
this problem are defined by the following scaling forms, valid for large $\tau$:
\bea
P(Q,\tau)\sim \rme^{-\tau h(Q/\tau)} \nn  \\
Z(\lambda) = \big\la \rme^{-\lambda Q} \big \ra =\int_{-\infty}^\infty
d Q\, \rme^{-\lambda
  Q}~P(Q,\tau) \sim  g(\lambda)~\rme^{ \tau\mu (\lambda)}. 
\eea
We will refer to $h(q)$ and $\mu(\lambda)$ as the LDF and CGF
respectively. For large $\tau$ the term $g(\lambda)$ is a correction
to the CGF and can normally be ignored in the saddle-point calculation
which relates $h(q)$ and $\mu(\lambda)$. The saddle-point calculation gives:
\bea
h(q)=-\bigl[\mu(\lambda^*) + q \lambda^*\bigr],~~~~ ~~\mu'(\lambda^*)=-q~.
\eea 
While the  LDF and the CGF are normally related by
Legendre transformations, there are  several examples where this relation
is known to break down. 
This happens if the function $g(\lambda)$ has
singularities in the region of the saddle-point integration.  
Interestingly, in such cases, the CGF might still satisfy the fluctuation
symmetry relation,  while the LDF does not \cite{visco06,farago:2002,vanZon:2003}. 
Thus we note that if one is interested in the LDF then it is in
general important to calculate both $\mu(\lambda)$  and the leading correction
term 
$g(\lambda)$. Of course $\mu(\lambda)$ is itself of interest since it
contains important information on current and current-noise properties
and relations between response functions.

The aim of this paper is to present a formalism to obtain
$\mu(\lambda)$ as well as $g(\lambda)$ for the problem of heat
conduction across a harmonic chain connected to white-noise Langevin
heat baths. We use the linearity of the problem and show that the
problem of finding the generating function $Z(\lambda)$ reduces to
performing multi-dimensional Gaussian integrations. We are able to
find a closed form expression for $\mu(\lambda)$, as given by
\eref{mu(lambda)}, in terms of the phonon-transmission function, a
well known quantity in the study of heat conduction in harmonic
systems. Finding $g(\lambda)$ is more difficult but we are able to
also express it, as given by \eref{expression-of-g(lambda)}, in terms
of appropriate phonon Green's functions.  For the case of a single
free Brownian particle we can use our approach to explicitly obtain
both $\mu(\lambda)$ and $g(\lambda)$ and for this case we recover the
results of Visco \cite{visco06}, obtained by solving the Fokker-Planck
equation.

The paper is organized as follows. In Sec.~(\ref{sec:model}) we define the
model that we study, make some general remarks on the problem of computing
the generating function for heat, and briefly explain our method. In
Sec.~(\ref{sec:cgf}) we give the calculation of the CGF $\mu(\lambda)$ while in
Sec.~(\ref{sec:correction}) we give the calculation of the correction term
$g(\lambda)$. The example of a single Brownian particle, 
for which explicit results for $\mu(\lambda)$ and $g(\lambda)$ can
be obtained, are considered in Sec.~(\ref{sec:examples}). Finally we discuss
our results in Sec.~(\ref{sec:discussions}).

\section{Model and general considerations}
\label{sec:model}
 
We consider a one-dimensional chain of $N$ particles with harmonic
interactions and described by the 
 Hamiltonian:
\begin{equation}
\mathcal{H}=\sum_{l=1}^N \frac{1}{2}m_l  v_l^2  + \frac{1}{2}\sum_{l=1}^{N}\sum_{m=1}^{N}~\bm\Phi_{lm}x_lx_m, \label{Ham}
\end{equation}
where $x_l$, $v_l$ and $m_l$ are, respectively, the displacement about
the equilibrium position, velocity and mass of $l$th particle and the
matrix $\bm{\Phi}$ represents the force matrix of the system.  For the
moment we assume that at least one site of the chain is pinned so that
the centre of mass attains a steady state distribution.  However,
later we show that the results are also valid for a free harmonic
chain.  The particles $1$ and $N$ at the two ends --- which we refer
as left (L) and right (R) respectively --- are coupled to white noise
Langevin heat reservoirs at two different temperatures $T_L$ and $T_R$
respectively.  The system, described by the variables
${X}^T=(x_1,x_2,\dots,x_N)$ and ${V}^T=(v_1,v_2,\dots,v_N)$, evolves
according to the following equations of motion:
\begin{equation}
\dot{{X}}={V},\quad
{\bm M} \dot{{V}}=-\bm{\Phi} X  -\bm{\gamma} V + \eta (t),
\label{Langevin}
\end{equation}
where $\bm{M}=\text{diag} (m_1,m_2,\dots,m_N)$ is the mass matrix, the
dissipation matrix $\bm{\gamma}$ has matrix elements ${\bm
  \gamma}_{i,j}=\delta_{i,j}(\delta_{i,1}\gamma_L
+\delta_{i,N}\gamma_R)$ and the noise vector ${\eta}$ has elements
${\eta}_i(t)= \delta_{i,1}\eta_L(t) +\delta_{i,N}\eta_R(t)$. The
variables $\eta_L(t),\eta_R(t)$ are zero-mean Gaussian white noises
with correlations given by:
\begin{equation}
\bigl\langle \eta_\alpha(t) \eta_{\alpha'}(t')\bigr\rangle = 2 
\delta_{\alpha,\alpha'} ~d_\alpha ~\delta(t-t')~,~~~{\rm  where}~~ d_{\alpha} =
\gamma_{\alpha}T_{\alpha}~, ~~~~\alpha,\alpha' = L,R~. ~~~
\end{equation}
and we have set the Boltzmann constant $k_B=1$.

Since the equations of motion \eref{Langevin} are linear and the
noise vector ${\eta}$ is Gaussian, the probability distribution
function of the phase space variables $U^T=(X^T,V^T)$ in the
nonequilibrium steady state is a Gaussian with mean $\la U\ra=0$ and
with covariance matrix $\lim_{t\rightarrow\infty}\langle {U}
{U}^T\rangle$.  We denote the nonequilibrium steady state distribution
by $P_\text{SS}(U)$.  The covariance matrix of the 
ordered harmonic chain was obtained exactly in \cite{RLL67}. For
mass-disordered systems the covariance matrix can be expressed in terms of
phonon Green's functions \cite{DR06,casher:1971}~. 
The quantity of our interest here is the total amount
of heat, $Q$, flowing from one of the reservoirs ---say the 
left (L) --- into the system in a given time duration $\tau$,
in the nonequilibrium steady state.  This is given by
\begin{equation}
  Q=\int_0^\tau  \bigl[\eta_L(t) - \gamma_L v_1(t)\bigr] v_1(t)\; d
  t,
\label{heat}
\end{equation}
where $v_1(t)$ evolves according to \eref{Langevin}, with the
initial condition at $t=0$ drawn from the nonequilibrium steady state
distribution.  Clearly, $Q$ is a fluctuating quantity whose value
depends on the initial conditions $U_0=U(t=0)$ and the noise
trajectory $\{{ \eta}(t): 0\le t \le \tau\}$ during any particular
realization.  Let $P(Q,\tau)$ denote the probability distribution of
$Q$ and let $Z(\lambda)= \bigl\langle \rme^{-\lambda Q}\bigr\rangle$ be
the corresponding characteristic function, where $\la...\ra$ denotes
an average over initial configurations as well as over different
paths.

It is useful to consider the restricted characteristic function
$Z(\lambda,U,\tau|U_0)=\bigl \langle \rme^{-\lambda
  Q}\bigr\rangle_{{U}_0,{U}}$ where the expectation is taken over all
trajectories of the system that evolve from a given initial
configuration ${U}_0$ to a given final configuration ${U}$ in time
$\tau$.  As shown in \ref{Fokker-Planck}, this satisfies a
Fokker-Planck-type equation: 
\bea \partial_\tau Z(\lambda,U,\tau|U_0)
= \mathcal{L}_\lambda Z(\lambda,U,\tau|U_0)~, \eea 
with the initial
condition $Z(\lambda,U,0|U_0)= \delta(U-U_0)$.  The solution of this
can formally be written down in the eigenbases of the Fokker-Planck
operator $\mathcal{L}_\lambda$, and the large $\tau$ behavior is
dominated by the term having the largest eigenvalue $\mu(\lambda)$,
i.e.,
\begin{equation}
  Z(\lambda,U,\tau |U_0) 
\sim
  \chi({U}_0,\lambda)\Psi({U},\lambda)\,
  \exp[\tau\mu(\lambda)]
  \label{characteristic.1}
\end{equation}
where $\Psi(U,\lambda)$ is the eigenfunction corresponding to
the largest eigenvalue, {\emph{i.e.}}, $\mathcal{L}_\lambda
\Psi({U},\lambda)=\mu(\lambda)\Psi({U},\lambda)$, and
$\chi({U}_0,\lambda)$ is the projection of the initial state onto
the eigenstate corresponding to the eigenvalue $\mu(\lambda)$.
We note that for $\lambda=0$, $Z(\lambda=0,U,\tau|U_0)$ is just the
phase space distribution at time $\tau$. Hence the existence of a unique
nonequilibrium steady state, which has been proved for this system
\cite{casher:1971},  requires that $Z(\lambda=0,U,\tau \to
\infty|U_0)=P_\text{SS}(U)$ and this implies that   $\mu(0)=0$, $\chi({U}_0,0)=1$
and $\Psi({U},0)=P_\text{SS}(U)$ ~.
Using \eref{characteristic.1}, and the fact that $P_\text{SS}(U)=\Psi(U,0)$,  we
get for large $\tau$:
\bea
Z(\lambda) = \int dU_0 \int dU \Psi(U_0,0) Z(\lambda,U,\tau|U_0)    \sim 
g(\lambda)\, \exp\bigl[\tau\mu(\lambda)\bigr],
\label{characteristic.2} \\
{\rm where~} g(\lambda) = \int dU_0  ~ \Psi({U}_0,0)
~\chi({U}_0,\lambda)~ \int d{U}~ \Psi({U},\lambda)~. \nn
\eea
Note that $g(0)=1$. 
As discussed in the introduction the large deviation function 
$h(q=Q/\tau)=-\lim_{\tau\rightarrow\infty} \ln P(Q,\tau)/\tau$
is given by the the Legendre transformation, 
\begin{equation}
h(q)=-\Bigl[\mu(\lambda^*)+\lambda^* q\Bigr],
\end{equation}
with $\lambda^*(q)$ implicitly given by the saddle point equation
$\mu'(\lambda^*)=-q$. The above relation holds {\emph{provided}} that
$g(\lambda)$ is analytic along the real $\lambda$ in the region
$[0,\lambda^*]$, so that $g(\lambda)$ can be neglected in the
saddle-point calculation as a subleading contribution and the contour
of integration can be deformed smoothly through the saddle point
$\lambda^*$.  On the other hand, if $g(\lambda)$ possesses any
singularity in the region $[0,\lambda^*]$, then the contour of the
integration cannot be deformed smoothly through the saddle point
$\lambda^*$, and one needs to include the singular part of
$g(\lambda)$ in the saddle point calculation.

The calculation of the LDF thus requires one to compute the CGF
$\mu(\lambda)$ and the leading correction $g(\lambda)$. From the above
discussion we see that these can be obtained from the largest
eigenvalue and eigenvector of an appropriate Fokker-Planck operator
for this system.  This is however very difficult in most cases,
including for the model studied here.  However the linearity of the
dynamics and the Gaussian nature of the noise in the present problem
allow the computation of $\mu(\lambda)$ and $g(\lambda)$ using a
different approach. The basic idea we use is that the variable of
interest $Q$ is a quadratic function of the initial phase-space
configuration $U_0$ and the noise-trajectories $\{\eta(t):~ 0 \leq t
\leq \tau\}$, both of which are Gaussian distributed variables.  Hence
the problem of computing $\big\la \rme^{-\lambda Q} \big\ra$ reduces to
one of doing a multi-variate Gaussian integration. In the following
sections we present the details.

We make some remarks on the symmetry property of the CGF. In
general, if the operator $\mathcal{L}_\lambda$ and its adjoint
$\mathcal{L}_\lambda^\dagger$ possess the symmetry
$\mathcal{L}_\lambda^\dagger = \mathcal{L}_{a-\lambda}$, then it 
immediately follows that $\mu(\lambda)=\mu(a-\lambda)$.
Even if another operator $\mathcal{L}'_\lambda$ ---
which is related to $\mathcal{L}_\lambda$ by a similarity transformation
--- possesses the symmetry ${\mathcal{L}'}_\lambda^\dagger =
{\mathcal{L}'}_{a-\lambda}$, then also $\mu(\lambda)$ has the above
symmetry.  There are some examples of systems with Markovian dynamics
where the evolution operator satisfies this property but this does not
seem to be the case for the model discussed here. 
In fact, even for the simplest case of a single free Brownian particle 
 connected to two heat reservoirs~\cite{visco06},
the Fokker-Planck operator does not possesses the above mentioned
symmetry. 
In this case however, the Fokker-Planck operator 
can be transformed to a Hermitian operator of a  quantum
harmonic oscillator where the potential remains invariant under
$\lambda\rightarrow (\Delta\beta-\lambda)$. 
We are not aware of such a transformation for a system having more
than one particle  and hence the symmetry of $\mu(\lambda)$ 
is a non-trivial one.

\section{Calculation of CGF $\mu(\lambda)$}
\label{sec:cgf}
Before giving the details of the calculation we 
first state our main result of this section for the CGF which is: 
 \begin{equation}
  \mu(\lambda)=-\frac{1}{4\pi}\int_{-\infty}^\infty d\omega\,
  \ln \Bigl[1+\mathcal{T}(\omega) T_L T_R
\,\lambda(\Delta\beta-\lambda)\Bigr],
\label{mu(lambda)}
\end{equation}
where $\Delta\beta=T^{-1}_R-T^{-1}_L$ and
\begin{eqnarray}
\label{T(omega)}
&\mathcal{T}(\omega)=&4\gamma_L\gamma_R
\omega^2\bm{G}^+_{1,N}(\omega)\bm{G}^-_{1,N}(\omega),\\
\label{G}
\text{with}\quad&\bm{G}^\pm(\omega)=&\Bigl[\bm{\Phi}
-\omega^2\bm{M}\pm i\omega\bm{\gamma}\Bigr]^{-1}~,
\end{eqnarray}
Note that ${\bm G}^-(\omega)={\bm G}^+(-\omega)={{\bm
    G}^+}^*(\omega)$~($^*$ denotes complex conjugate).  Using
\eref{mu(lambda)}, one can immediately verify the expression for
the average energy current,
\begin{equation}
\lim_{\tau\rightarrow\infty}\frac{\langle Q\rangle}{\tau}=-\mu'(0)=
\frac{T_L-T_R}{4\pi}
\int_{-\infty}^\infty d\omega\, \mathcal{T}(\omega).
\end{equation}
which was obtained previously~\cite{dhar:2001, casher:1971,
  rubin:1971, o'connor:1974}.  From \eref{mu(lambda)}, we  note that
$\mu(0)=0$ as required.  It is also evident that \eref{mu(lambda)}
possesses the symmetry $\mu(\lambda)=\mu(\Delta\beta-\lambda)$.  
We also note that (\ref{mu(lambda)}) agrees with the classical limit of
the result obtained in \cite{keijidhar} for a quantum chain.

We now give the details of the derivation of the expression
\eref{mu(lambda)}.  We solve the Langevin equations of motion by
Fourier transforms. Let us define the finite-time Fourier transforms
and their inverses as follows:
\begin{eqnarray}
\{\widetilde{X}(\omega_n),\widetilde{V}(\omega_n),\widetilde{\eta}(\omega_n)
\}  = \frac{1}{\tau} \int_0^\tau ~dt~ \{{X}(t),{V}(t),{\eta}(t) \}
\exp(-i \omega_n t)\, \nn \\
\{{X}(t),{V}(t),{\eta}(t) \}=
\sum_{n=-\infty}^\infty
\{\widetilde{X}(\omega_n),\widetilde{V}(\omega_n),\widetilde{\eta}(\omega_n) 
\} ~ \exp(i \omega_n t)\, \nn \\
{\rm with} ~\omega_n = \frac{2\pi n}{\tau}~. 
\end{eqnarray}
The Gaussian noise configurations represented by $\{{
  \eta}(t):0<t<\tau\}$, can now equivalently be described in the
frequency domain, by the infinite sequence $\{\widetilde{
  \eta}(\omega_n):n = -\infty, \ldots, -1, 0,\ldots,\infty\}$ of
Gaussian random variables having the correlations
\begin{equation}
  \bigl\langle \widetilde{\eta}_\alpha(\omega)
  \widetilde{\eta}_{\alpha'}(\omega')\bigr\rangle =2
  \delta_{\alpha,\alpha'}\frac{\gamma_\alpha T_\alpha}{\tau}\,
\delta[\omega+\omega'],\quad \text{with}\quad
  \alpha,\alpha'=L,R~.
\end{equation}
Henceforth for convenience we will drop the subscript $n$ from $\omega_n$.
Taking the Fourier transform of \eref{Langevin} gives the velocity vector in
frequency domain as
\begin{eqnarray}
\widetilde{{V}}(\omega) &=&  i \omega \bm G^+ ~ \widetilde{\eta} +
\frac{1}{\tau} {\bm G}^+ \big(~ \bm\Phi \Delta {X}_\tau - i \omega {\bm M} \Delta
{V}_\tau ~\big ), \label{vomga1} \\ 
{\text{where}} ~& &\Delta {X}_\tau = {X}(\tau)-{X}(0),~\Delta {V}_\tau =
{V}(\tau)-{V}(0)~,\nonumber  
\end{eqnarray}
and the Green's function matrix ${\bm G}^+(\omega)$ is given by  \eref{G}. 
Since $\widetilde{\eta} \sim 1/\tau^{1/2}$ and $\Delta X_\tau,~\Delta V_\tau$
have finite variances for large $\tau$, it 
follows that  the second term in (\ref{vomga1}) is
$O(1/\tau^{1/2})$ smaller than the first. As will be shown in the next section
they contribute to the correction term $g(\lambda)$. In this section we
focus on computing  $\mu(\lambda)$ and the first term in (\ref{vomga1}) 
then gives:
\begin{equation}
\widetilde{v}_1(\omega) \sim i\omega \bigl[{G}^+_{1,1}(\omega)
  \,\widetilde{\eta}_L(\omega) +
{G}^+_{1,N}(\omega)\,
  \widetilde{\eta}_R(\omega)
 \bigr]~.
\label{v_1(omega)}
\end{equation}
In terms of the Fourier transform, the expression in \eref{heat} for the
heat transfer becomes
\begin{equation}
  Q(\tau)=\tau \sum_{n=-\infty}^\infty \widetilde{q}_n
\quad\text{with}\quad \widetilde{q}_n=
\widetilde{\eta}_L(\omega) \widetilde{v}_1(-\omega)
 - \gamma_L \widetilde{v}_1(\omega)\widetilde{v}_1(-\omega).
\label{heat in omega}
\end{equation}
For a chain with at least one pinned site it is easily seen from
\eref{v_1(omega)} that $\widetilde{v}_1(0)=0$ and hence the heat
transfer $\widetilde{q}_0$ through the zero-th mode vanishes in
\eref{heat in omega}.  This is related to the fact that for a
pinned system there is no zero-frequency mode available for
transporting energy.  Substituting \eref{v_1(omega)} in
\eref{heat in omega} we write $ Q(\tau)=\tau \sum_{n=1}^\infty
(\widetilde{q}_n + \widetilde{q}_{-n})$ and using the fact
$\widetilde{\eta}_\alpha(-\omega)=\widetilde{\eta}^*_\alpha(\omega)$
we get:
\begin{equation}
Z(\lambda) =\big \la \rme^{-\lambda Q} \big \ra \sim \prod_{n=1}^\infty
\left\langle \rme^{-\lambda\,\tau\, {\xi}_n^T \bm{A}_n {\xi}_n^* }
\right\rangle,
\label{characteristic.3}
\end{equation}
\begin{eqnarray}
\fl\text{where}\quad
{\xi}_n&=&\Bigl( \widetilde{\eta}_L(\omega),  
\widetilde{\eta}_R(\omega)\Bigr)^T \nn \\
\fl\text{and}\quad\bm{A}_n &=&
\left(\begin{array}{cc}
2 \gamma_R \omega^2 |{ G}^+_{1,N}|^2  
&\quad-i \omega {G}^-_{1,N} - 2 \gamma_L \omega^2
{G}^+_{1,1}{G}^-_{1,N} \\ 
i \omega {G}^+_{1,N} - 2 \gamma_L \omega^2
{G}^-_{1,1}{G}^+_{1,N} 
& -2 \gamma_L \omega^2|{G}^+_{1,N}|^2 
\end{array}\right)~,
\label{A_n}
\end{eqnarray}
where in obtaining the $(1,1)^{\rm th}$ element above we have made use of the
identity 
\begin{equation}
 -\text{Im} {G}^+_{1,1}(\omega)
=\omega ~\left[
\gamma_L   |{G}^+_{1,1}(\omega)|^2  
  +\gamma_R   |{G}^+_{1,N}(\omega)|^2\right]~,
\label{identity.1}
\end{equation}
which can be proved as follows. From the definition of the Green's
function in \eref{G} we have $[\bm{G}^+]^{-1} - [\bm{G}^-]^{-1} = 2
i \omega \bm{\gamma}$. Multiplying both sides by $\bm{G}^+$ from the
left, and by $\bi{G}^-$ from the right we get $\bm{G}^- - \bm{G}^+ = 2
i \omega \bm{G}^+~ \bm{\gamma} \bm{G}^-$.  The $(1,1)^{\text{th}}$
element of this leads to \eref{identity.1}~.  For each $n$, the
average in \eref{characteristic.3} is evaluated with respect to
the Gaussian distribution
\begin{equation}
\fl p({\xi}_n)= \frac{1}{\pi^2 \text{det}\bi{D}}
\exp\left(-{ \xi}_n^T \bi{D}^{-1} {\xi}_n^*\right)~,
\quad\text{with} ~~\bi{D} = 
\text{diag}\left(\frac{2d_L}{\tau},\frac{2d_R}{\tau}\right)~. 
\label{G-dist}
\end{equation}
Hence we get (see \ref{Gaussian integral})
\begin{eqnarray}
\left\langle \rme^{-\lambda\tau {\xi}_n^T \bm{A}_n {\xi}_n^*
}\right\rangle &=& \biggl [\text{det}(\bm{I} +\lambda \tau \bm{D A}_n
  )\biggr ]^{-1} ~,\\ &=& \Bigl[1+\mathcal{T}(\omega) T_L T_R
  \,\lambda(\Delta\beta-\lambda)\Bigr]^{-1}, \label{characteristic.4}
\end{eqnarray}
where $\mathcal{T}(\omega)$ is given by \eref{T(omega)}.  
Using (\ref{characteristic.3},\ref{characteristic.4}) and noting that 
in the $\tau\rightarrow\infty$ limit we can replace the summation over $n$ 
by an integral over $\omega$ we obtain our final result in
\eref{mu(lambda)}.

We remark that, the result \eref{mu(lambda)} is in fact valid for both
pinned as well as unpinned cases.  For the unpinned harmonic system
the centre of mass coordinate does not reach a steady state and it is
useful to separate out this degree of freedom. Let us define new
relative coordinates $y_l=x_l-x_N,~l=1,2,\ldots,N-1$. The facts that
the unpinned system has translational symmetry and that ${\bm \Phi}$
is a symmetric matrix imply the relation $\sum_{l=1,N}{\Phi}_{l,j}
=\sum_{j=1,N}{ \Phi}_{l,j} = 0$. Using this we get from
\eref{Langevin}: \bea \dot{y}_l &=& v_l-v_N,
~~~{l=1,2,\ldots,N-1} \label{yeq1} \\ m_l \dot{v}_l&=&
-\sum_{j=1}^{N-1} {\Phi}_{l,j} y_j -{\gamma}_{l,l} v_l +
{\eta}_l(t),~~~~~~~{l=1,2,\ldots,N}~. \label{yeq2}
\label{modLangevin}
\eea
The relative coordinates $y_l,~l=1,2,\ldots,N-1$ attain steady state with
finite variance and it follows then from \eref{yeq1} that 
$\tilde{v}_l(\omega=0)\sim \tilde{v}_N(\omega=0)$  up to order
$O(1/\tau)$. Then from \eref{yeq2} we get $\widetilde{v}_l(0)\sim
[\widetilde{\eta}_L(0) + \widetilde{\eta}_R(0)]/[\gamma_L+\gamma_R]$
for all $l$. We  use this to compute the 
$\widetilde{q}_0\not=0$ term in \eref{heat in omega}. 
With ${\xi}_0=\bigl(\widetilde{\eta}_L(0),  
\widetilde{\eta}_R(0)\bigr)^T$, we get 
\begin{equation}
\widetilde{q}_0  \sim  \frac{1}{2}{\xi}_0^T \bi{A}_0 {\xi}_0 \quad
{\rm where}~~ \bi{A}_0 = \left(\begin{array}{rr} 
\frac{2\gamma_R}{(\gamma_L+\gamma_R)^2} 
&\quad-\frac{\gamma_L-\gamma_R}{(\gamma_L+\gamma_R)^2} \\
-\frac{\gamma_L-\gamma_R}{(\gamma_L+\gamma_R)^2} 
&-\frac{2\gamma_L}{(\gamma_L+\gamma_R)^2}  \nn
\end{array}\right)~.
\end{equation}
The noise $\xi_0$ has the Gaussian distribution 
\begin{equation}
 p({\xi}_0)= \frac{1}{2\pi \sqrt{\text{det}\bm{D}}}
\exp\left(-\frac{1}{2}{\xi}_0^T \bm{D}^{-1} {\xi}_0\right)
\label{GD-k0}
\end{equation}
and hence we get
\begin{equation}
\left\langle \rme^{-\lambda\tau \widetilde{q}_0}\right\rangle 
\sim \Bigl[\det(\bm{I}+\lambda\tau \bm{D}\bm{A}_0)\Bigr]^{-1/2} \nn
=\Bigl[1+\mathcal{T}(0) T_L T_R
  \lambda(\Delta\beta-\lambda)\Bigr]^{-1/2} 
\end{equation}
where $\mathcal{T}(0)=4\gamma_L\gamma_R/(\gamma_L+\gamma_R)^2$ is
precisely what one obtains by taking $\mathcal{T}(\omega \to 0)$ in
\eref{T(omega)} for the unpinned system.  This follows from the
fact that in this case ${G}^+_{1N}|_{\omega\rightarrow 0} \sim
{1}/{i\omega(\gamma_L+\gamma_R)}$, a result which we now prove: For
any matrix ${\bm A}$ let $ {\bm A}^{(i,j)}_{(k,l)}$ denote the
submatrix of ${\bm A}$ that occurs between the $(i,j)^{\rm th}$ and
the $(k,l)^{\rm th}$ elements. Also let ${\bm B}=-{\bm M}
\omega^2+{\bm \Phi}+i \omega {\bm \gamma}$ . Then we have ${
  G}^+_{1,N}= (-1)^{N+1}\det {\bm B}^{(1,2)}_{(N-1,N)} \Big/ \det {\bm
  B}$. Taylor expanding the determinants about $\omega=0$ we obtain $
\det {\bm B}^{(1,2)}_{(N-1,N)}= \det {\bm \Phi}^{(1,2)}_{(N-1,N)} +
O(\omega)$ and $ \det{\bm B}=i \omega \gamma_L \det {\bm
  \Phi}^{(2,2)}_{(N,N)} +i \omega \gamma_R \det{\bm
  \Phi}^{(1,1)}_{(N-1,N-1)}+ O(\omega^2) $, where we have used $\det
\bm{\Phi}=0$ that follows from the property $\sum_{l=1,N}{
  \Phi}_{l,j} =\sum_{j=1,N}{ \Phi}_{l,j} = 0$. Using the latter
property it is easy to show that $ \det{\bm
  \Phi}^{(2,2)}_{(N,N)}=\det{\bm \Phi}^{(1,1)}_{(N-1,N-1)}=
(-1)^{N-1}\det {\bm \Phi}^{(1,2)}_{(N-1,N)}$.  Hence we get the
desired result.

\section{Calculation of $g(\lambda)$}
\label{sec:correction}

We turn now to the more difficult problem of calculating $g(\lambda)$ which
requires one to keep the second term in \eref{vomga1} and perform the
average  over initial conditions, in addition to the noise average. We recall
that the heat transfer is given by (\ref{heat in omega}) where the
velocity $\tilde{v}_1$  can be obtained from the following exact 
solution for $\widetilde{X},\widetilde{V}$: 
\begin{eqnarray}
\widetilde{X} = \bm{G}^+ ~\widetilde{\eta} - \frac{1}{\tau} {\bm G}^+ ~\bigl
          ( {i \omega \bm{M}\Delta X_\tau+{\bm \gamma}\Delta X_\tau}
          +\bm{M} \Delta V_\tau 
            \bigr)~, \nn \\
\widetilde{{V}} =  i \omega \bm G^+ ~ \widetilde{\eta} +
\frac{1}{\tau} {\bm G}^+ \big(~ \bm\Phi \Delta {X}_\tau - i \omega
          {\bm M} \Delta {V}_\tau ~\big )~, \label{xomga1} \\  
{\rm where}~\Delta {X}_\tau = {X}(\tau)-{X}(0),~\Delta {V}_\tau =
{V}(\tau)-{V}(0)~.\nonumber  
\end{eqnarray}
To obtain $Z(\lambda) = \la \rme^{-\lambda Q} \ra$ we need to average over
both noise $\tilde{\eta}$ and over the initial steady state
distribution of $U_0$. We note that the solution in \eref{xomga1}
contains $U^T(\tau)=(X^T(\tau),V^T(\tau))$ and this has to be
expressed in terms of $\tilde{\eta}$ and $U_0$. While this can be done
we follow a different strategy which is more convenient.

It is  easier to calculate the restricted generating function
\begin{eqnarray}
Z(\lambda,U,\tau |U_0)= \Biggl \langle {{e}}^{{-\lambda Q } } \delta
\bigl ({U} - {U}(\tau) \bigr )\Biggr \rangle_{U_0}
~, \label{Rgenf}
\end{eqnarray}
where the average is performed only over noise and for a given initial
condition $U_0$.  We can then obtain $Z(\lambda)= \int dU \int dU_0
P_\text{SS}(U_0) Z(\lambda,U,\tau|U_0)$.  The steps of the calculation
now are as follows.
 We first note that, because of the $\delta$-function constraint in
 (\ref{Rgenf}),we can obtain $Q$ as  a quadratic function of the variables
 $\tilde{\eta},U,U_0$.  
This follows  from (\ref{heat in omega}) by using the following
expression for $\tilde{v}_1(\omega)$ obtained from (\ref{xomga1})
by replacing  $U(\tau)$ by $U$: 
\bea
\fl \tilde{v}_1(\omega) &=& i \omega [~{G}^+_{11}~\tilde{\eta}_L +
  {G}^+_{1N}~\tilde{\eta}_R~] +\frac{1}{\tau} \sum_{l=1,N} [ (\bm{G}^+
  \bm{\Phi})_{1l} \Delta X_l -i \omega  (\bm{G}^+
  \bm{M})_{1l} \Delta V_l ] \nn \\
\fl &=&  i \omega [~{G}^+_{11}~\tilde{\eta}_L +
  {G}^+_{1N}~\tilde{\eta}_R~] + \frac{1}{\tau} F_1^T \Delta U~,  \label{v1ex}
\eea
where
\bea
\fl\Delta X &=& X-X(0),~\Delta V= V-V(0),~ \Delta U= U-U(0),~~~{\rm
  and} \nn \\
\fl  F_1^T&=& \Bigl(
[\bm{G}^+\bm{\Phi}]_{1,1},[\bm{G}^+\bm{\Phi}]_{1,2},\dots,[\bm{G}^+\bm{\Phi}]_{1,N},
-i\omega [\bm{G}^+\bm{M}]_{1,1},-i \omega
[\bm{G}^+\bm{M}]_{1,2},\dots,-i \omega [\bm{G}^+\bm{M}]_{1,N} 
\Bigr)~.~~~~~~\label{F1def}
\eea

 We replace the $\delta$-function in (\ref{Rgenf}) by the integral
 representations: $\delta(U - U(\tau)) = \int
 {d^{2N}{\sigma}}/{(2\pi)^{2N}}~\rme^{i \sigma^T (U - U(\tau))}$ 
 where $\sigma^T=(\sigma_1,\sigma_2,...\sigma_{2N}) $.  We then have:
\begin{equation}
Z(\lambda, U, \tau|,U_0) = \int \frac{d^{2N}\sigma}{(2\pi)^{2N}} 
~\rme^{i \sigma^T U } \bigl \langle \rme^{E(\tau)}\bigr \rangle_{U,U_0}
\label{Z-2}
\end{equation}
where $E(\tau)=-\lambda Q -i \sigma^T U(\tau)  $.

 We need an expression for $U^T(\tau)=(X^T(\tau),~V^T(\tau))$ which we now
  obtain. We note that since we are using a Fourier-series
  representation for $X(t),V(t)$, the correct value at time $\tau$ is
  obtained from the Fourier-series by setting
  $t=\tau-\epsilon,~\epsilon >0$ and taking the limit $\epsilon \to
  0$. Hence we obtain: 
\begin{eqnarray}
U^T=(~X^T(\tau),~V^T(\tau)~) = \lim_{\epsilon \to
  0}\sum_{n=-\infty}^\infty \left(~ 
\widetilde{X}^T (\omega_n),~\widetilde{V}^T(\omega_n) ~\right) ~ \rme^{-i \omega_n
  \epsilon}~. \nonumber
\end{eqnarray}
For large $\tau$ we note that $(1/\tau)~ \sum_n {\bm G}^+(\omega_n)
\rme^{-i \omega_n \epsilon} =0 $ which follows from converting the
summation into a integral and noting that all the poles of
$G^+(\omega_n)$ lie in the upper half plane.  Hence we get:
\begin{eqnarray}
 U^T(\tau) =(~X^T(\tau),~V^T(\tau)~)&= \sum_{n=-\infty}^\infty ~\left(~ [\bm{G}^+
\widetilde{\eta}]^T ,~ i \omega [\bm{G}^+ \widetilde{\eta}]^T ~\right)~\rme^{-i
  \omega_n \epsilon}\nn \\
 &=\sum_{n=-\infty}^\infty ~ \rme^{-i   \omega_n \epsilon} ~ 
\Bigl[ F_2^T \tilde{\eta}_L+ F_3^T
\tilde{\eta}_R\Bigr]~,   \label{utau} 
\end{eqnarray}
where
\begin{eqnarray}
 F_2^T&= \Bigl(
{G}^+_{1,1},~{G}^+_{2,1},~\dots,{G}^+_{N,1},~
i\omega {G}^+_{1,1},~i \omega {G}^+_{2,1},\dots,i \omega {G}^+_{N,1}~ \Bigr), \label{F2def}\\
 F_3^T&= \Bigl(
{G}^+_{1,N},~{G}^+_{2,N},\dots,{G}^+_{N,N},~
i\omega {G}^+_{1,N},~i \omega{G}^+_{2,N},\dots,i\omega {G}^+_{N,N}\Bigr)~.\label{F3def}
\label{vtau}
\end{eqnarray}
 Hence we get $E(\tau)$ in \eref{Z-2} as
\begin{equation}
E(\tau)=-\lambda Q - i \sum_{n=-\infty}^{\infty} \rme^{-i\omega_n \epsilon}
\Bigl[ \sigma^T F_2 ~ \tilde{\eta}_L +\sigma^T F_3 ~\tilde{\eta}_R\Bigr]~ .
\label{sk}
\end{equation}
After using the full expression for $\widetilde{v}_1(\omega)$ from
(\ref{v1ex}) to evaluate $Q$ in (\ref{heat in omega}), we obtain 
$E(\tau) = s_0 + \sum_{n=1}^{\infty} (s_n + s_{-n})$ where $(s_n +
s_{-n})$  has the following quadratic form
\begin{eqnarray}
s_n+s_{-n} &=&-\lambda\tau\xi_n^T \bi{A}_n \xi_n^* + \xi^T_n \alpha_n +
\alpha^{T}_{-n}\xi_n^* +
\frac{2\lambda\gamma_L}{\tau}|f_n|^2~, \label{somega}
\end{eqnarray}
where the matrix $\bi{A}_n$ is given by 
\eref{A_n} and
\bea
  \alpha_n &=& \lambda\left(\begin{array}{c}
2i\gamma_L  \omega G^+_{1,1} -1 \\
2i\gamma_L  \omega G^+_{1,N}
\end{array}\right) F_1^\dagger \Delta U
- i \rme^{-i \omega \epsilon} 
\left(\begin{array}{c}
  F_2^T \sigma \\
 F_3^T \sigma
\end{array}\right)~,  \label{alpdef} \\
f_n &=& F_1^T \Delta U \label{fdef}, \eea where $F_1,F_2$ and $F_3$
are given by \eref{F1def},\eref{F2def} and \eref{F3def} respectively.

Similarly, one can express $s_0$ as
\begin{eqnarray}
s_0 &=&-\frac{\lambda\tau}{2} \xi_0^T \bm{A}_0 \xi_0 + \alpha^T_0\xi_0
+ \frac{\lambda\gamma_L}{\tau}f_0^2, \label{s0omega}
\end{eqnarray}
where $\bi{A}_0, \alpha_0$, $\xi_0$ and $f_0$ are all real.

We now first evaluate averages with respect to
the Gaussian distribution given in \eref{G-dist} for $n\ne0$ and
with distribution given in \eref{GD-k0} for $n=0$.  We get
\bea
\fl \bigl \langle \rme^{ (s_n+s_{-n})} \bigr \rangle_{{U},{U}_0} &=&
\frac{1}{\text{det}(\bm{I} +\lambda
  \tau \bm{D A}_n )} \exp \left[ ~\alpha^T_{-n}(\bm{D}^{-1} + 
    \lambda\tau\bm{A}_n)^{-1}\alpha_n + \frac{2\lambda\gamma_L}{\tau}|f_n|^2
    ~\right]  
\quad ~ n \ne 0, \label{A1} \nn \\
\fl \bigl \langle \rme^{ s_0} \bigr \rangle_{U,U_0} &=&
\frac{1}{\sqrt{\text{det}(\bm{I} +\lambda \tau \bm{D A}_0)}} \exp \left[
  \frac{1}{2} \alpha^T_{0} (\bm{D}^{-1} + \lambda\tau\bm{A}_0)^{-1}     \alpha_0 
+ \frac{\lambda\gamma_L}{\tau}|f_0|^2~\right]~. \label{A2}
\eea
Hence we get 
\begin{eqnarray}\fl
\bigl \langle \rme^{E(\tau)}\bigr \rangle_{{U},{U}_0}=
\exp\left(-\frac{1}{2}\sum_{n=-\infty}^\infty\ln \biggl
         [\text{det}(\bm{I} +\lambda \tau \bm{D A}_n )\biggr
         ]\right)\nn\\
\times\,
\exp \left(\sum_{n=-\infty}^{\infty} 
\Biggl[\frac{1}{2}
\alpha^T_{-n}(\bm{D}^{-1} +
    \lambda\tau\bm{A}_n)^{-1}\alpha_n
+\frac{\lambda\gamma_L}{\tau}
|f_n|^2
\Biggr] 
\right) 
\label{BB} 
\end{eqnarray} 
After evaluating the required matrix inverse and determinant, we take the
large $\tau$ limit to replace all the summations over $n$ by integrations
over $\omega$. This  yields 
\begin{equation*}
\fl \bigl \langle \rme^E \bigr \rangle_{U,U_0} \sim \rme^{\tau
  \mu(\lambda)}\,
\exp\left(\int_{-\infty}^{\infty}
\frac{d\omega}{2\pi}\Biggl[
\frac{d_Ld_R\,\alpha^{T}(-\omega)\bm{C}(\omega)\alpha(\omega)}{1+T_LT_R\lambda(\Delta
  \beta -\lambda) \mathcal{T}(\omega)}
+ \lambda\gamma_L |f(\omega)|^2
\Biggr]\right),
\end{equation*}
where $\mu(\lambda)$ is given by \eref{mu(lambda)} and
\begin{equation}\fl
\bm{C} (\omega) =
\left(\begin{array}{cc}
\frac{1}{d_R}  -4 \lambda\gamma_L \omega^2|{G}^+_{1,N}|^2 
&\quad 4 \lambda\gamma_L \omega^2
{G}^+_{1,1}{G}^-_{1,N}  + 2i \lambda\omega {G}^-_{1,N} \\ 
4\lambda \gamma_L \omega^2
{G}^-_{1,1}{G}^+_{1,N}  -2i\lambda \omega {G}^+_{1,N} 
&\frac{1}{d_L} + 4\lambda \gamma_R \omega^2 |{ G}^+_{1,N}|^2  
\end{array}\right).
\label{C_n}
\end{equation}

We see from (\ref{alpdef},\ref{fdef}) that $\alpha(\omega),~f(\omega)$ are
linear in $\Delta  U$ and $\sigma$. After some algebraic manipulations and use
of the identity 
(\ref{identity.1}) we then arrive at the following compact expression for
$\bigl \langle \rme^E \bigr \rangle_{U,U_0}$:
\begin{equation}\fl
\bigl \langle \rme^E \bigr \rangle_{U,U_0}\sim \rme^{\tau \mu(\lambda)}
\exp\Bigl ( -\frac{1}{2}{\sigma}^T \bm{H}_1~{\sigma} + i \Delta U^T \bm{H}_2~{\sigma}
+\frac{1}{2} \Delta U^T \bm{H}_3~\Delta U\Bigr ), \label{B2}
\end{equation}
where
\begin{eqnarray}
\label{H1}
&\fl \bm{H}_1 (\lambda) = \frac{1}{2} (~\bm{I}_1+\bm{I}_1^T~)~,
\\ &\fl\text{with}\qquad\bm{I}_1 (\lambda)=
\frac{d_Ld_R}{\pi}\int_{-\infty}^{\infty} d\omega~ \frac{C_{1,1}F_2
  F_2^{\dagger} + C_{1,2}F_3 F_2^{\dagger} +C_{2,1}F_2 F_3^{\dagger} +
  C_{2,2}F_3 F_3^{\dagger}} {1+T_LT_R\lambda(\Delta \beta -\lambda)
  \mathcal{T}(\omega)}~,\nn
\end{eqnarray}
\begin{equation}
\fl \bm{H}_2(\lambda) =
\lim_{\epsilon\rightarrow 0}
\frac{\lambda}{\pi}
\int_{-\infty}^{\infty}
d\omega\,
 \rme^{i \omega \epsilon}\,
\frac{d_L (1-2i\omega \gamma_L
    {\bi{G}^+_{1,1}} )F_1^* F_2^{\dagger} 
-2i\omega (\gamma_L + \lambda d_L)d_R\,\bi{G}^+_{1,N}F_1^* F_3^{\dagger}
}{1+T_LT_R\lambda(\Delta \beta -\lambda) \mathcal{T}(\omega)}~,
\label{H2}
\end{equation}
and 
\begin{eqnarray}
\label{H3}
&\fl \bm{H}_3(\lambda) =\frac{1}{2} (~\bm{I}_3+\bm{I}_3^T~)~,  \\
&\fl\text{with}\qquad
\bm{I}_3(\lambda) = \frac{\lambda \bigl ( \gamma_L + \lambda d_L
  \bigr)}{\pi}\int_{-\infty}^{\infty} d\omega~\frac{F_1F_1^{\dagger}
}{1+T_LT_R\lambda(\Delta \beta -\lambda) \mathcal{T}(\omega)}~.\nn
\end{eqnarray}

Finally to get $Z(\lambda,U,\tau|U_0)$ we substitute the
expression for $\bigl \langle \rme^E \bigr \rangle_{U,U_0}$ from
\eref{B2} into \eref{Z-2} and perform the Gaussian integration over
$\sigma$. This gives
\begin{eqnarray}
\fl Z(\lambda, U, \tau|U_0) \sim \frac{\rme^{\tau
    \mu(\lambda)}}{(2\pi)^{N}~\sqrt{\det \bm{H}_1}}~\rme^{\frac{1}{2}
  \Delta U^T \bm{H}_3~\Delta U}~~ \rme^{-\frac{1}{2}(U^T+\Delta
  U^T\bm{H}_2)\bm{H}_1^{-1}(U+\bm{H}_2^T\Delta U)}.\label{p_lamb}
\end{eqnarray}
Putting $\lambda=0$ in the above expression gives the steady state
distribution as
\begin{equation}
P_\text{SS}(U)= Z(0, U, \tau\rightarrow\infty|U_0) =
\frac{\exp\left(-\frac{1}{2} U^T \bm{H}_1^{-1}(0) U\right)}
{(2\pi)^{N}~\sqrt{\det \bm{H}_1(0)}}~. \label{Pss}
\end{equation}
From the long time solution in (\ref{utau}) it can be directly verified that
\begin{equation*}
\lim_{t\rightarrow\infty} \langle U U^T\rangle =\frac{1}{2\pi}
\int_{-\infty}^{\infty} d\omega~
\Bigl[d_L F_2 F_2^{\dagger} + d_RF_3 F_3^{\dagger}\Bigr]~,
\end{equation*}
and using the fact that  $\langle U U^T\rangle$ is real we see that the above
equals  $\bm{H}_1(0)$, consistent with (\ref{Pss})

Now, according \eref{characteristic.1}, the initial and the final
variables $U_0$ and $U$ in \eref{p_lamb} must factorize, which implies
\begin{math}
 \bigl[\bm{H}_3-\bm{H}_2 \bm{H}_1^{-1} \bm{H}_2^T -  
\bm{H}_1^{-1} \bm{H}_2^T\bigr]
+
\bigl[\bm{H}_3-\bm{H}_2 \bm{H}_1^{-1} \bm{H}_2^T -  
\bm{H}_2 \bm{H}_1^{-1}\bigr]^T=\bm{0}.
\end{math}
Since $\bm{H}_1$ and $\bm{H}_3$ are symmetric matrices, the above
condition can be expressed as
\begin{equation}
 \bm{H}_3-\bm{H}_2 \bm{H}_1^{-1} \bm{H}_2^T - \bm{H}_1^{-1}
   \bm{H}_2^T=\bm{0}~,
\label{condition-H}
\end{equation}
Using this \eref{p_lamb} gives
\begin{equation}
\fl Z(\lambda, U, \tau|,U_0) \sim \frac{\rme^{\tau
    \mu(\lambda)}}{(2\pi)^{N}~\sqrt{\det \bm{H}_1(\lambda)}}\,
\exp\left(-\frac{1}{2} U^T \bm{L}_1(\lambda) U\right)\,
\exp\left(-\frac{1}{2} U^T_0 \bm{L}_2(\lambda) U_0\right)~.
\end{equation}
This means that we can make the following identifications for
$\Psi(U,\lambda),~\chi(U_0,\lambda)$:
\bea
\label{wave function}
 &&\Psi(U,\lambda) = \frac{1}{(2\pi)^{N}~\sqrt{\det \bm{H}_1(\lambda)}}~
\exp\left(-\frac{1}{2} U^T \bm{L}_1(\lambda) U\right)~, \\
\label{projection}
 &&\chi(U_0,\lambda) = \exp\left(-\frac{1}{2} U^T_0 \bm{L}_2(\lambda)
U_0\right)~, \\
\fl {\rm where} && \bm{L}_1(\lambda) =\bm{H}_1^{-1}+ \bm{H}_1^{-1} \bm{H}_2^T \\
\fl \text{and} && \bm{L}_2(\lambda)=\bm{H}_2 \bm{H}_1^{-1} \bm{H}_2^T 
-\bm{H}_3 =-\bm{H}_1^{-1} \bm{H}_2^T~.
\eea
Thus we have obtained the left and right eigenvectors of the Fokker-Planck
operator $\mathcal{L}_\lambda$ corresponding to the eigenvalue
$\mu(\lambda)$. It can be seen that the orthonormality condition $\int dU
\chi(U,\lambda)\Psi(U,\lambda) =1$ is satisfied. 

We obtain $Z(\lambda)$ by integrating $Z(\lambda,U,\tau|U_0)$ over $U$
and then averaging over the 
initial condition $U_0$ with respect to the steady state distribution
$P_\text{SS}(U_0)$. This then  gives our final expression for the correction
to the CGF:
\begin{equation}
g(\lambda) =\Bigl(\det \bm{H}_1(\lambda) \det \bm{H}_1 (0)
 \det\bm{L}_1(\lambda)  \det[\bm{H}_1^{-1}(0) +  \bm{L}_2(\lambda)]
\Bigr)^{-1/2}.
\label{expression-of-g(lambda)}
\end{equation}
Since $\bm{L}_1(0)= \bm{H}_1^{-1}(0)$ and $\bm{L}_2(0)= \bm{0}$, it is
verified that $g(0)=1$.

\section{Example of Single Brownian particle}
\label{sec:examples}

The Langevin equation for a single Brownian particle is given by :
\begin{equation}
m\dot{v} = -(\gamma_L+\gamma_R)v + \eta_L(t) + \eta_R(t), \label{eqm-s-bpart}
\end{equation}
where $v$ is the velocity of the particle and $m$ is it's mass. Here
we consider the velocity of the particle not the position since
velocity $v$ of the particle will have a normalized steady state
distribution whereas position will not have and the heat transfer $Q$,
in which we are interested, does not depend on position. For single
Brownian particle the matrix defined in (\ref{G}) becomes a
complex number: ${\bm G}^+(\omega)= 1/(-m\omega^2 +i\omega \gamma)$
where $\gamma=\gamma_L+\gamma_R$. Following all the steps described in
the last section one can easily arrive at the expression \eref{B2}
where, $\mu(\lambda)$ is given in \eref{mu(lambda)} and $\bm{H}$'s
are given in \eref{H1}, \eref{H2} and \eref{H3}. In this case one
can carry out the integrations present in the expressions of all these
quantities. The expression for phonon transmission coefficient is
obtained from \eref{T(omega)} and given by $\mathcal{T}(\omega)=4
\gamma_L\gamma_R [m^2\omega^2 + (\gamma_L+\gamma_R)^2]^{-1}$. We use
this form in \eref{mu(lambda)} to evaluate the integral and get
\begin{equation}
  \mu(\lambda)=\frac{\gamma_L+\gamma_R}{2m}
  \left[1-\sqrt{1+\frac{4\gamma_L\gamma_R}{(\gamma_L+\gamma_R)^2}
 {T_LT_R\lambda(\Delta\beta-\lambda)}}\;
  \right].
\end{equation}
This is the result obtained in \cite{visco06}.  Similarly using ${\bm
  G}^+(\omega)= 1/(-m\omega^2 +i\omega \gamma)$ and the above form for
$\mathcal{T}(\omega)$ we evaluate $\bm{H}$'s given by \eref{H1},
\eref{H2} and \eref{H3} to get
\begin{eqnarray}
 &\bm{H}_1 = \frac{d_L +d_R}{m  \sqrt{ \gamma^2 + a^2 } }, \nonumber \\
 &\bm{H}_2 = \frac{\lambda d_L + \frac{1}{2}(\gamma -\sqrt{\gamma^2 + a^2 })}{ \sqrt{ \gamma^2 + a^2 }}, \nonumber \\
  &\bm{H}_3 = \frac{m \lambda}{  \sqrt{ \gamma^2 + a^2 } }\Bigl( \gamma_L + \lambda d_L \Bigr), \nonumber \\
\fl  {\text{where,}} &a =\sqrt{4 d_L d_R \lambda (\Delta \beta - \lambda )}. 
\end{eqnarray}
It is easily verified that $\bm{H}$s' for the single Brownian particle
satisfy the relation \eref{condition-H}, {\emph{i.e.}} 
$(\bm{H}_2)^2 + \bm{H}_2 - \bm{H}_1\bm{H}_3=0$.
Now using the expression \eref{expression-of-g(lambda)} we obtain
\begin{equation}
g(\lambda)
=2\,\sqrt{\frac{\gamma\sqrt{\gamma^2 + 4 d_L d_R\lambda(\Delta \beta -\lambda)}}{\bigl (\gamma+\sqrt{\gamma^2 + 4 d_L d_R\lambda(\Delta \beta -\lambda)}~\bigr )^2 - 4\lambda^2 d_L^2}}~,
\end{equation}
which also agrees with the result in \cite{visco06}.

\section{Discussions}
\label{sec:discussions}
We have presented a formalism to calculate the CGF $\mu(\lambda)$ and it's
correction $g(\lambda)$ for heat transport across a  harmonic chain connected
to white-noise Langevin reservoirs. The formula for $\mu(\lambda)$ is
expressed as an integral over frequencies, with the integrand depending
explicitly on the phonon  transmission function $\mathcal{T}(\omega)$. The
expression for $g(\lambda)$ 
is in terms of integrals involving an appropriate  phonon Green's function. 
We have illustrated the  usefulness of the 
formalism by calculating $\mu(\lambda)$ and $g(\lambda)$ for a single Brownian
particle for which case all  integrals can be performed explicitly. 
For systems with more number of particles, the function $\mathcal{T}(\omega)$
can easily be  obtained analytically for an ordered harmonic chain and
numerically for disordered harmonic chains. Hence our formalism can be used to
numerically compute $\mu(\lambda)$ and $g(\lambda)$ with high accuracy.
A knowledge of these functions would enable one to check the validity of the
fluctuation symmetry for  the large deviation function. We note  that
$\mu(\lambda)$ itself is a useful quantity, containing information on current
moments in the nonequilibrium state. We show that it always satisfies the
fluctuation symmetry relation.  
Finally we have pointed out that $\mu(\lambda)$ can, in general,  be shown to
be  the largest eigenvalue of a Fokker-Planck type operator
($\mathcal{L}_\lambda$ for our problem). Using our formalism we not only
obtain this eigenvalue but also the corresponding left and right eigenvectors.

The present approach has recently been generalized to the problem of computing
$\mu(\lambda)$ for the case of heat conduction across arbitrary harmonic
networks \cite{SD2010}. The problem of calculating $g(\lambda)$ in such cases
and also the extension of the present formalism to quantum systems are interesting
open problems.

\appendix

\section{The Fokker-Planck equation}
\label{Fokker-Planck}
Let $P(Q,U,t|U_0)$ denotes the probability distribution of heat flow $Q$ in duration $\tau$ given the initial and final configuration $U$ and $U_0$ respectively. 
The distribution $P(Q,U,t|U_0)$ satisfies the following Fokker-Planck equation
\begin{eqnarray}
\fl \frac{\partial P}{\partial t} &=& \Bigl[-\sum_{l=1}^{2N} \frac{\partial}{\partial U_l} \frac{\langle \Delta U_l \rangle}{\Delta t}  
- \frac{\partial}{\partial Q}\frac{\langle \Delta Q \rangle}{\Delta t} 
+\frac{1}{2} \sum_{l=1}^{2N}\sum_{m=1}^{2N}\frac{\partial^2}{\partial U_l\partial U_m} \frac{\langle \Delta U_l \Delta U_m \rangle}{\Delta t} \nn \\
\fl &+&\sum_{l=1}^{2N}\frac{\partial^2}{\partial U_l \partial Q}\frac{\langle \Delta U_l \Delta Q \rangle}{\Delta t} 
+\frac{1}{2}\frac{\partial^2}{\partial Q^2}\frac{\langle \Delta Q^2 \rangle}{\Delta t} \Big ] P(Q,U,t|U_0)~;~~{\rm{with}}~\Delta t \to 0~, \label{FP1}
\end{eqnarray}
where the moments are calculated using the Langevin equations (\ref{Langevin}) and heat equation given in (\ref{heat}). After calculating the moments we get 
\begin{equation}
\frac{\partial P}{\partial t} =  {\mathcal{L}}_Q~ P(Q,U,t|U_0) 
\end{equation}
where
\begin{eqnarray}
{\mathcal{L}}_Q = &&\sum_{l=1}^N\frac{1}{m_l}\Big[ \frac{\partial \mathcal{H}}{\partial x_l}\frac{\partial}{\partial v_l}
- \frac{\partial \mathcal{H}}{\partial v_l}\frac{\partial}{\partial x_l}\Big]\nn \\ 
&&+\frac{\gamma_L}{m_1}\frac{\partial}{\partial v_1}v_1 
+\frac{\gamma_R}{m_N}\frac{\partial}{\partial v_N}v_N  +\left(\gamma_L v_1^2-\frac{\gamma_L T_L}{m_1}\right) \frac{\partial}{\partial Q} \nn \\ 
&&+ \frac{\gamma_L T_L}{m_1^2}\frac{\partial^2}{\partial v_1^2} 
+\frac{\gamma_R T_R}{m_N^2}\frac{\partial^2}{\partial v_N^2} + \gamma_L T_L v_1^2 \frac{\partial^2}{\partial Q^2} 
+\frac{2\gamma_L T_L}{m_1}\frac{\partial^2}{\partial v_1 \partial Q}v_1. \label{FP}
\end{eqnarray}
The corresponding Fokker-Planck equation for the restricted characteristic function $Z(\lambda,U,t|U_0)= \int_{-\infty}^{\infty}dQ \rme^{-\lambda Q} P(Q,U,t|U_0) $ is obtained by 
multiplying both sides of the above equation by $\rme^{-\lambda Q}$ and then integrating with respect to $Q$. We get
\begin{equation}
 \frac{\partial }{\partial \tau}{Z}(\lambda,U,\tau|U_0) =
    {\mathcal{L}}_{\lambda} ~ {Z}(\lambda,U,\tau|U_0) 
\end{equation}
where
\begin{eqnarray}
 {\mathcal{L}}_{\lambda}=&\sum_{l=1}^N\frac{1}{m_l}\Big[ \frac{\partial \mathcal{H}}{\partial x_l}\frac{\partial}{\partial v_l}
- \frac{\partial \mathcal{H}}{\partial v_l}\frac{\partial}{\partial x_l}\Big]\nonumber\\
&+\frac{\gamma_L}{m_1}\frac{\partial}{\partial v_1}v_1 
+\frac{\gamma_R}{m_N}\frac{\partial}{\partial v_N}v_N  +\left(\gamma_L v_1^2-\frac{d_L}{m_1}\right)~\lambda \nonumber \\
 &+ \frac{d_L}{m_1^2}\frac{\partial^2}{\partial v_1^2} 
+\frac{d_R}{m_N^2}\frac{\partial^2}{\partial v_N^2} + d_L v_1^2 \lambda^2 
+\frac{2d_L}{m_1} \lambda\frac{\partial}{\partial v_1}v_1 .~~~~
\end{eqnarray}

\section{Multidimensional Gaussian integral of complex variables}
\label{Gaussian integral}
For ease of reference we give the following result for complex Gaussian
integrals. 
\begin{equation}
\int d^n Z \exp(-Z^T \bm{A} Z^* + Z^T B+ 
C^T Z^*) = \frac{\pi^{n}}{\text{det}\bm{A}}\,\exp({C^T\bm{A}^{-1} B})
\label{comGauss}
\end{equation}
where $Z$ is a $n$-dimensional complex vector, $\bm{A}$ is a
Hermitian matrix~ and $B,C$ are arbitrary complex $n$-dimensional vectors.  
The integration $\int d^n Z$ denotes the real integrations $\int d^n X
d^n Y$ with the substitution $Z=X+iY$,  $X$ and $Y$ being real
$n$-dimensional vectors.

\noindent {\bf Proof:}
Let $\bm{K}$ be the unitary matrix such that ${\bm A}=\bm{K}{\bm
  D}\bm{K}^{\dagger}$ where $\bm{D}$ is a diagonal matrix with all real  
diagonal elements. Now with the following transformations 
\begin{eqnarray}
\widetilde{Z} = \bm{K}^TZ~, 
\widetilde{B} &=& \bm{K}^{\dagger}B~,~\widetilde{C} = \bm{K}^TC
\end{eqnarray}
we can write the complex Gaussian integration in the  form
\begin{eqnarray}
&&\int_{\infty}^{\infty} d^n \widetilde{X}d^n \widetilde{Y} \exp(-\widetilde{Z}^T\bm{D}\widetilde{Z}^* +\widetilde{Z}^T\widetilde{B} + \widetilde{C}^T\widetilde{Z}^*). 
\end{eqnarray}
Since $\bm{D}$ is diagonal, the above is a product of $2n$ uncoupled Gaussian
integrations. Performing the integrations we get:  
\begin{eqnarray}
\fl \prod_{i=1}^n \frac{\pi}{D_i} \exp(\frac{\widetilde{B}_i\widetilde{C}_i}{D_i}) 
=\frac{\pi^n}{\det \bm{A}} \exp(\widetilde{C}^T\bm{D}^{-1}\widetilde{B}) 
= \frac{\pi^n}{\det \bm{A}}\exp({C}^T\bm{A}^{-1}{B})~~,
\end{eqnarray} 
which completes the proof.


\section*{References}
\begin{thebibliography}{20}
 

 \bibitem{ECM93}
D. J. Evans, E. G. D. Cohen, and G. P. Morriss, Phys. Rev. Lett. {\bf 71},
2401 (1993).
\bibitem{ES94}  D. J. Evans and D. J. Searles, Phys. Rev. E {\bf 50}, 1645
  (1994).
\bibitem{GC95} G. Gallavotti and E.G.D. Cohen, Phys. Rev. Lett. {\bf 74}, 2694
(1995).
 \bibitem{Jarz97} C. Jarzynski, Phys. Rev. Lett. {\bf 78}, 2690 (1997).
\bibitem{Kur98} J. Kurchan, J. Phys. A: Math. Gen. {\bf 31},
  3719 (1998). 
\bibitem{Crooks99} G. E. Crooks,  Phys. Rev. E {\bf 60}, 2721 (1999).
\bibitem{LS99} J. L. Lebowitz and H. Spohn, J. Stat. Phys. {\bf 95}, 333
  (1999). 

\bibitem{HS01}  T. Hatano and S. Sasa, Phys. Rev. Lett. {\bf 86}, 3463 (2001).
\bibitem{ND04} O. Narayan and A. Dhar, J. Phys. A: Math. Gen. {\bf 37} 63
  (2004). 
\bibitem{Seifert05}  U. Seifert, Phys. rev. lett. {\bf 95}, 040602, (2005).
\bibitem{JW04} C. Jarzynski and D. K. Wojcik,  Phys. Rev. Lett. {\bf 92},
  230602 (2004).  
\bibitem{BD04}  T. Bodineau and B. Derrida,  Phys. Rev. Lett. {\bf 92},
  180601 (2004).
\bibitem{ED04}  C. Enaud and B. Derrida, J. Stat. Phys. {\bf 114}, 537 (2004).
\bibitem{DDR04}B. Derrida, B. Doucot and P.-E. Roche  J. Stat. Phys. {\bf 115},
  717-748 (2004). 
\bibitem{DL98} B. Derrida and J.L. Lebowitz,  Phys. Rev. Lett. {\bf 80}, 209
  (1998). 
\bibitem{visco06} P. Visco, J. Stat. Mech. P06006 (2006).

\bibitem{keijidhar} K. Saito and A. Dhar, Phys. Rev. Lett. {\bf 99}, 180601 (2007).

\bibitem{AG06}D. Andrieux and P. Gaspard, Phys. Rev. E 74, 011906 (2006).

\bibitem{lacoste08} D. Lacoste, A. W.C. Lau, and K. Mallick, Phys. Rev. E {\bf
  78}, 011915 (2008). 


\bibitem{Wang02} G. M. Wang \etal, Phys. Rev. Lett. {\bf 89}, 050601 (2002).

\bibitem{Carberry04}D. M. Carberry \etal, Phys. Rev. Lett. {\bf 92}, 140601 (2004).

\bibitem{FM04}  K. Feitosa and N. Menon,  {\emph ibid} {\bf 92}, 164301
  (2004). 
\bibitem{Goldburg01} W. I. Goldburg \etal, Phys. Rev. Lett. {\bf 87}, 245502
  (2001).   
\bibitem{Douarche06} F. Douarche \etal, {\emph  ibid} {\bf 97}, 140603 (2006).
  
\bibitem{Liphardt02} J. Liphardt \etal, Science {\bf 296}, 1832 (2002).
 
\bibitem{Collin05} D. Collin \etal, Nature {\bf 437}, 231 (2005).

\bibitem{Solano10} J. R. Gomez-Solano \etal, Europhys Lett. {\bf 89} 60003
  (2010). 

\bibitem{MS08} S. Majumdar and A. K. Sood, Phys. Rev. Lett. {\bf 101}, 078301
  (2008).  

\bibitem{touchette:2009} H. Touchette, Phys. Rep. {\bf 478}, 1 (2009).


\bibitem{gomez} A. Gomez-Marin  and J. M. Sancho , Phys. Rev. E {\bf 73}, 045101(R),(2006).
\bibitem{Wijland:2006} F. van Wijland, Phys. Rev. E {\bf 74}, 063101 (2006).



\bibitem{farago:2002} J. Farago, J. Stat. Phys., {\bf 107},  781
  (2002).

\bibitem{vanZon:2003} R. van Zon and E. G. D.  Cohen,
  Phys. Rev. Lett. {\bf 91}, 110601 (2003); Phys. Rev. E {\bf 69},
  056121 (2004).


\bibitem{RLL67} Z. Rieder, J. L. Lebowitz and E. Lieb, J. Math. Phys.
{\bf 8}, 1073 (1967).

\bibitem{DR06} A. Dhar and D. Roy, J. Stat Phys. {\bf 125}, 801 (2006).

\bibitem{casher:1971} A. Casher and J.L. Lebowitz, J. Math. Phys. {\bf
    12}, 1701 (1971).

\bibitem{dhar:2001} A. Dhar, Phys. Rev. Lett. {\bf 86}, 5882 (2001).


\bibitem{rubin:1971} R. J. Rubin and W. L. Greer, J. Math. Phys. {\bf
    12}, 1686 (1971).

\bibitem{o'connor:1974} A. J. O'Connor and J. L. Lebowitz,
  J. Math. Phys. {\bf 15}, 692 (1974).

\bibitem{SD2010} K. Saito and A. Dhar, arXiv:1012.0622.

\end{thebibliography}
\end{document}